\def \cm{~\rm{cm}}
\def \s{~\rm{s}}
\def \km{~\rm{km}}

\def \K{~\rm{K}}
\def \g{~\rm{g}}

\def \AU{~\rm{AU}}

\def \yr{~\rm{yr}}
\def \year{~\rm{year}}

\documentclass[12pt,preprint]{aastex}

\shorttitle{Circumbinary Disks} \shortauthors{Akashi \& Soker}

\begin{document}

\title{A MODEL FOR THE FORMATION OF LARGE CIRCUMBINARY DISKS AROUND POST AGB STARS}

\author{ Muhammad Akashi and Noam Soker,\altaffilmark{1} }

\altaffiltext{1}{Department of Physics,
Technion$-$Israel Institute of Technology, Haifa 32000, Israel;
akashi@physics.technion.ac.il; soker@physics.technion.ac.il. }

\begin{abstract}
We propose that the large, radius of $\sim 10^3 \AU$, circumbinary rotating disks observed around some
post-asymptotic giant branch (post-AGB) binary stars are formed from slow AGB wind material that is
pushed back to the center of the nebula by wide jets.
We perform 2D-axisymmetrical numerical simulations of fast and wide jets that
interact with the previously ejected slow AGB wind.
In each system there are two oppositely launched jets, but we use the symmetry of the problem and
simulate only one jet.
A large circularization-flow (vortex) is formed to the side of the jet which together with the thermal
pressure of the shocked jet material accelerate cold slow-wind gas back to the center from
distances of $\sim 10^3-10^4 \AU$.
We find for the parameters we use that up to $\sim 10^{-3} M_\odot$ is back-flowing to the center.
We conjecture that the orbital angular momentum of the disk material results from the
non-axisymmetric structure of jets launched by an orbiting companion.
This conjecture will have to be tested with 3D numerical codes.
\end{abstract}

\section{INTRODUCTION}
\label{sec:intro}

Among the many riddles related to the transition from the asymptotic giant
branch (AGB) to the planetary nebula (PN) phase, prominent is the presence
of large circumbinary Keplerian disks around some post-AGB stars
(Waters et al. 1993, 1998; Jura et al. 1995, 1997, 2000;
Van Winckel et al. 1995, 1998; Jura \& Kahane 1999, Van Winckel 1999;
Bujarrabal et al. 2003, 2005; Dominik et al.\ 2003; Hinkle et al. 2007).
These post AGB stars (the primary) are found to have a close companion (the secondary)
at a typical orbital separation of $a \sim 1 \AU$, while the disk is much larger, up to
$\sim 2000 \AU$ (de Ruyter et al. 2006).
In most cases the companion is thought to be a low mass main sequence
star, but can also be a WD (e.g., Hinkle et al. 2007).
In some systems the large-disk resides in the center of a much larger bipolar nebula,
e.g., the Red Rectangle (Cohen et al.2004) and 89 Here (Bujarrabal et al. 2007).

The total angular momentum in these disks is small compared to that of the binary system
because the total mass
 in the disk is small $\la 10^{-2} M_\odot$
(Bujarrabal et al. 2003, 2005).
However, the specific angular momentum in the disk is about an order of magnitude larger
than that in the binary system. This is not easy to explain (Soker 2000).
One possible explanation is that the binary system exerts torque on the inner boundary of
the disk and transfers angular momentum to the disk via viscosity (Jura et al. 2002; Frankowski 2007).
It is not clear, however, if there is sufficient time for the viscosity to distribute angular
momentum in such large disks (see section 3).
This still leaves the question of how the material extends to distances of  $\sim 10-1000$
times the orbital separation without dispersing first. After all, most of the gas is expelled
from AGB and post-AGB stars.
Motivated by the large specific angular momentum in these disks, Soker (2000) suggested that the
dense matter in the equatorial plane is a dense slow equatorial flow rather than a Keplerian disk.
However, the disk in the Red Rectangle was found to be Keplerian (e.g.,
Bujarrabal et al. 2005), although it also has an outflowing component.

In this paper we propose that the circumbinary disk is formed by the interaction of jets
with the slow wind of the AGB or post-AGB primary star, that cause material to
fall from large distance back toward the center.
In this preliminary study we present the principles of the flow structure,
and don't try to fit specific cases.

\section{NUMERICAL SIMULATIONS}
\label{sec:numeric}

The simulations were performed using Virginia Hydrodynamics-I
(VH-1), a high-resolution multidimensional astrophysical hydrodynamics
code (Blondin et al. 1990; Stevens et al., 1992; Blondin 1994).
For gas temperatures above $10^4 \K$ we use the radiative cooling function for
solar abundances from Sutherland \& Dopita (1993; see Akashi et al. 2007),
while for $T \le 10^4 \K$ we use the radiative cooling time as given by Woitke et al.
(1996; their fig. 11).
For numerical reasons we set a minimum temperature at $200 \K$.
We use cylindrical (axisymmetrical) grid, namely, we simulate 3D flow with a 2D grid.
Only one quarter of the meridional plane is simulated, as the other three
quarters are symmetric to it.
There are 208 grids points in the $90^\circ$ azimuthal direction and 208 grids points in
the radial direction, with a cell size that increases with radius.

We show here the results for two models.
The first one has the following parameters.
At the beginning of each simulation, $t=0$, the grid is filled with the cold ($1000 \K$) slow
wind material having an outflow velocity of $v_1=5 \km \s^{-1}$.
We assume that 950 years before the beginning of the jet-launching phase
the slow wind mass loss rate increased by a factor of four.
At $t=0$ we take the density for $r \le 3 \times 10^{16} \cm$ to be that for a slow wind
mass loss rate of $\dot M_1 = 2 \times 10^{-5} M_\odot \yr^{-1}$, and for $r > 3 \times 10^{16} \cm$
the density corresponds to a mass loss rate of $\dot M_1 = 5 \times 10^{-6} M_\odot \yr^{-1}$.
At $t=0$ we start to inject a conical jet within a half opening angle of $\alpha=60^\circ$ and
a constant speed of $v_j=600 \km \s^{-1}$.
In recent years there are more indications for wide jets, in clusters of galaxies and
in PNs (Sternberg et al. 2007; Sternberg \& Soker 2008; Soker 2004),
and therefore our use of a wide jet is justified.
The inner boundary of the grid is at $r=10^{15} \cm$.
The jet is injected in a conical nozzle with its outlet at $r=7 \times 10^{15} \cm$.
The mass injection rate of one jet is $\dot M_j=10^{-7} M_\odot \yr^{-1}$,
injected uniformly within the angle $\alpha$.
In the second model we did not increase the mass loss rate of the slow wind prior
to the jet launching.

For numerical reasons we inject a very weak wind between the jet and
the equatorial plane during the jet-launching phase.
This outflow that must be incorporated to prevent numerical failure, prevents mass
from being accreted directly to the center.
For that reason, we examine the mass back-flow rate at larger distances.
Therefore, the flow close to the center, where the equatorial slow outflow can be
seen in the Figures, is not real. We rather expect an inflow there.

\begin{figure}
\resizebox{0.99\textwidth}{!}{\includegraphics{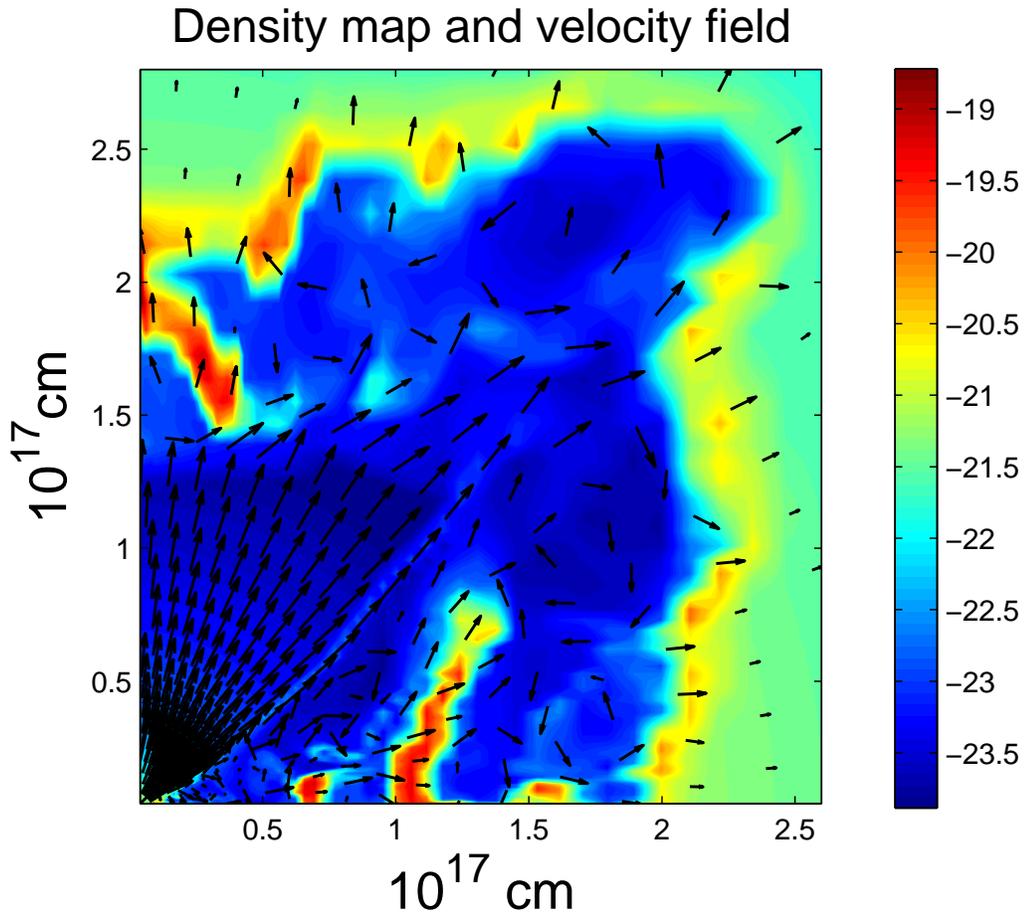}}
\caption{The density (scale on the right in $\log(\g \cm^{-3})$) and velocity plot in one quarter of the
meridional plane at $t=1900~$years after the onset of the jet, in the first model
(with increased slow wind mass loss rate 950 years before jet launching).
The arrows indicate the direction of the flow, and for three velocity ranges:
$v > 200 \km \s^{-1}$ (long arrow), $ 20 < v \le 200 \km \s^{-1}$ (medium arrows), and
$ v \le 20 \km \s^{-1}$ (short arrows).
The vertical axis is the symmetry axis, and the horizonal axis is in the equatorial plane.  }
\label{nebA1}
\end{figure}

\begin{figure}
\resizebox{0.99\textwidth}{!}{\includegraphics{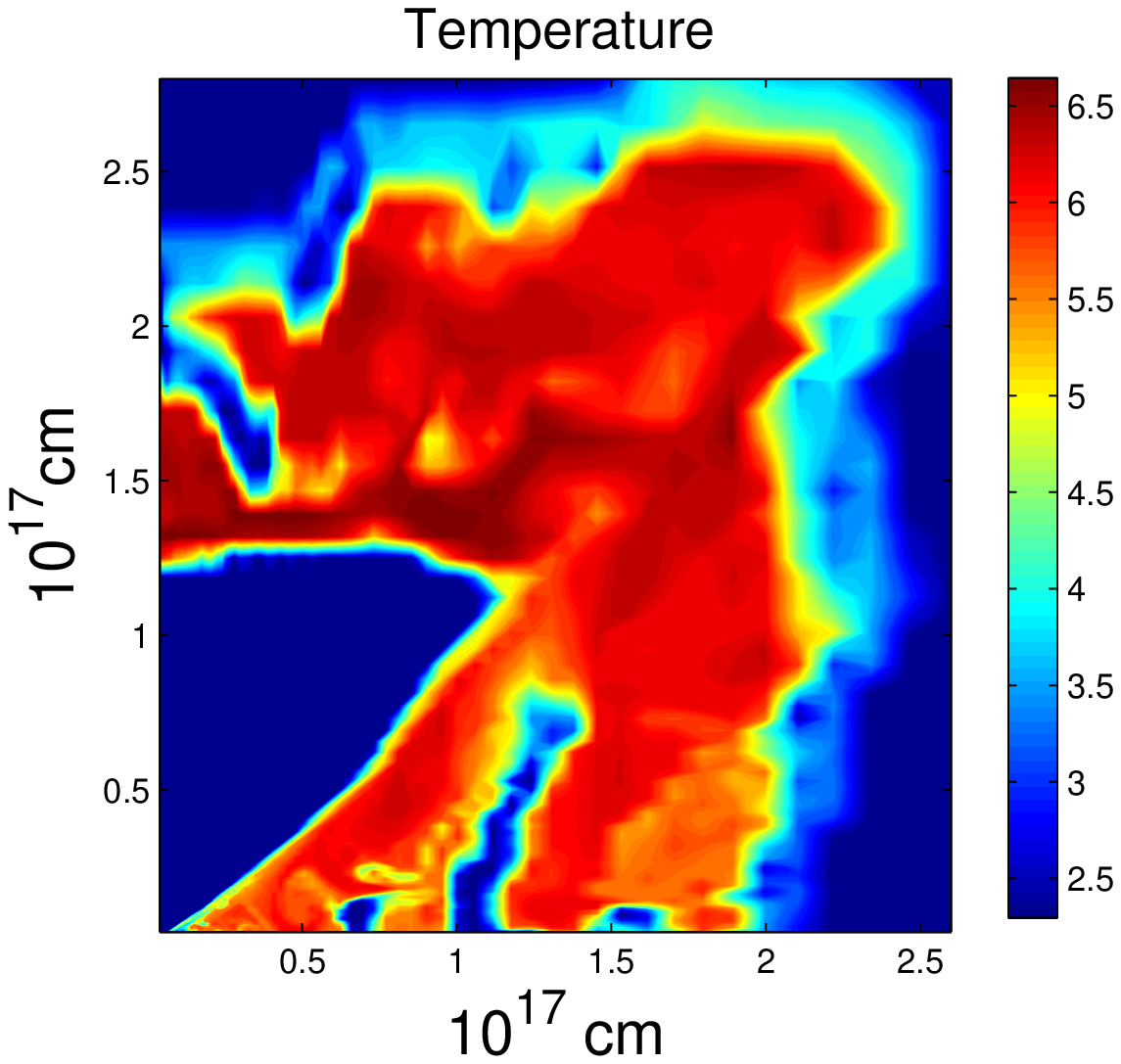}}
\caption{The temperature (scale on the right in $\log(\K)$)
map corresponding to the same run and time as in Figure \ref{nebA1}. }
\label{nebA2}
\end{figure}

In Figure \ref{nebA1} we show the entire nebula at $t=1900~$years, where
the colors represent the density and the arrows show the velocity.
The corresponding gas temperature is shown in Figure \ref{nebA2}.
The structural features developed along the symmetry axis are not real.
They may result form real physical instabilities, but their rate of growth and their
exact structure are not real, but rather due to numerical limitations of the 2D numerical
code. These numerical problems are limited to the vicinity of the symmetry axis.
As we are interested in features near the equatorial plane, we will ignore the
flow near the symmetry axis in this work, and postpone the discussion to the next paper.

Most relevant to us is the flow of the `cocoon' and the back-flow toward the center
that is formed later close to the equatorial plane.
The cocoon$-$a well known feature of expanding jets$-$is the slowly moving material
around the expanding jet, which is formed from the post-shock jet material and some ambient
matter (the slow-wind gas).
In our simulations the cocoon forms a low density large circulating flow to the sides
of the jets (in the axisymmetrical simulations the structure is a torus).
The center of this vortex is seen at $(x,z)=(1.6,1)\times 10^{17} \cm$,
where $x$ is in the equatorial plane and $z$ is along the symmetry axis.
The jets and cocoon form two large low density bubbles, one on each side of the equatorial plane
(in our simulation we only show one quarter of the meridional plane).
As can be clearly seen there is a dense shell around each bubble.
In a full 3D image we have formed a bipolar nebula.

\begin{figure}
\hskip -0. cm  
\resizebox{0.99\textwidth}{!}{\includegraphics{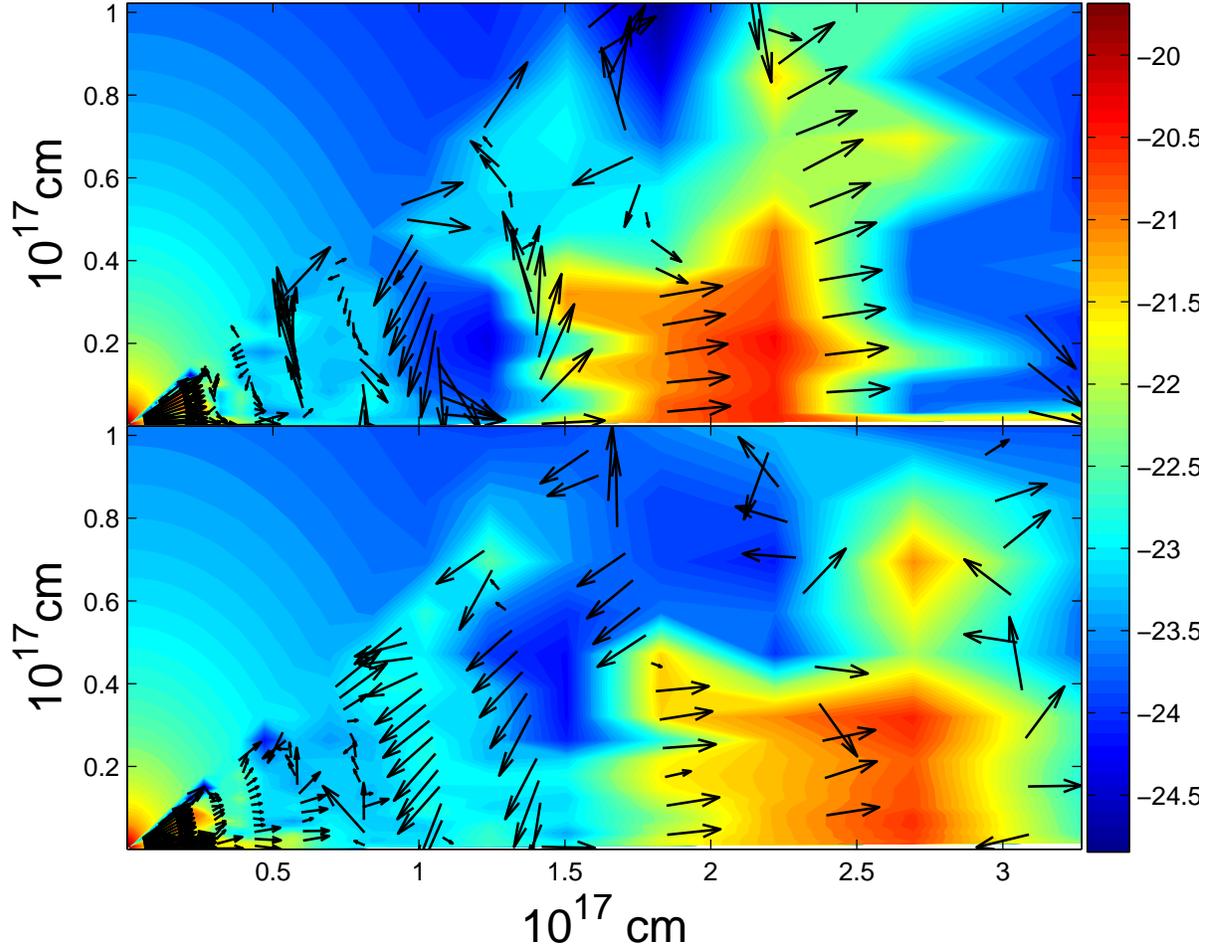}}
\caption{The inner region of the simulated grid for two late times of the first model:
$4800~$years (upper panel) and $5800~$years (lower panel).
Note the inflowing material.
Shown are the density (scale on the top in $\g \cm^{-3}$) and the velocity,
represented by arrows for three velocity ranges:
$10 < v \le 100 \km \s^{-1}$ (long arrow), $ 5 < v \le 10 \km \s^{-1}$ (medium arrows), and
$ v \le 5 \km \s^{-1}$ (short arrows).
The vertical axis is the symmetry axis, and the horizonal axis is in the equatorial plane.
Note that the dense equatorial outflow near the center, $r \la 2 \times 10^{16} \cm$, is introduced
to prevent numerical failure. In reality we expect an inflow there, resulting from the
back-flowing gas. }
\label{nebB1}
\end{figure}

As can be seen in Figures \ref{nebA1} and \ref{nebA2}, the vortex touches a cool,
$T \simeq 200-10^4 \K$, and dense material, seen as a stripe extending from
$(x,z)=(1.1,0)\times 10^{17} \cm$ to $(x,z)=(1.3,0.7)\times 10^{17} \cm$.
Later this vortex (the circularization flow)  and the thermal pressure of the hot bubble
push cool material back to the center, as can be seen in Figure \ref{nebB1}
that show the inner region of the grid at late times.
This is the basic result of this and other simulations we performed
with other parameters. Namely, that dense material can fall back toward the center
from very large distances of few$\times 10^{17} \cm$.

In the simulation presented here the mass of the back-flowing material increases
with time.
As stated, the total back-flowing mass is calculated by summing
the inflowing gas mass at regions away from the slow equatorial outflow close to
the center (this outflow was introduced for numerical reason).
At $t = 7000~$years in the first model the total inflowing mass reaches a values of
$M_{\rm back} \simeq 7 \times 10^{-4} M_\odot$.
For other parameters we get different masses, but of that order or smaller.

\begin{figure}
\hskip -0.0 cm
\resizebox{0.99\textwidth}{!}{\includegraphics{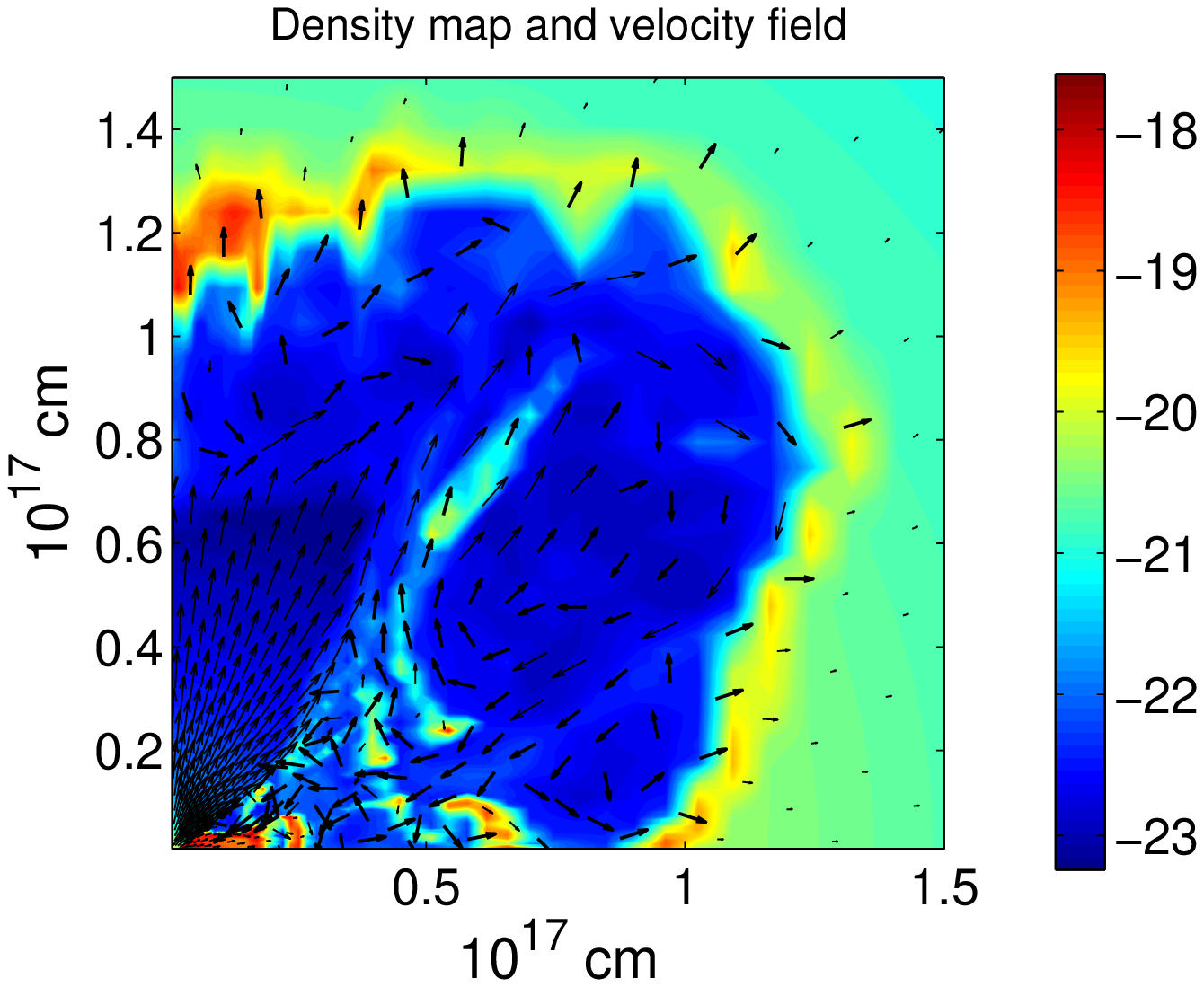}}
\caption{The density (scale on the right in $\g \cm^{-3}$) and velocity plot in one quarter of the
meridional plane at $t=1300~$years after the onset of the jet in the second model
(the model without increase in the slow wind mass loss rate).
The arrows indicate the direction of the flow, and for three velocity ranges:
$v > 200 \km \s^{-1}$ (long arrow), $ 20 < v \le 200 \km \s^{-1}$ (medium arrows), and
$ v \le 20 \km \s^{-1}$ (short arrows).
The vertical axis is the symmetry axis, and the horizonal axis is in the equatorial plane.
Note that the dense equatorial outflow near the center, $r \la 2 \times 10^{16} \cm$, is introduced
to prevent numerical failure. In reality we expect an inflow there, resulting from the
back-flowing gas. }
\label{neb4}
\end{figure}

The result of the second model (the one without increased slow wind mass loss rate)
at $t=1300~$year is shown in Figure \ref{neb4}.
The total (within the entire grid) back-flowing mass reaches values of
$M_{\rm back} \simeq 2-3 \times 10^{-4} M_\odot$.
The back-flowing mass within a radius of $6 \times 10^{16} \cm$ and $3 \times 10^{16} \cm$,
is $2 \times 10^{-4} M_\odot$ and $2 \times 10^{-5} M_\odot$, respectively.
This is because at that time most of the back-flowing mass resides in the dense blobs
seen in Figure \ref{neb4} in at distances of $4 \times 10^{16} \cm \la r \la 5 \times 10^{16} \cm$.

There are other features of the simulations that are less relevant to us.
We mention two of these.
As was proposed by Soker \& Rappaport (2000), the pressure of the bubbles compresses material
in the equatorial plane. The pressure is composed of both the thermal pressure of the
relatively hot gas inside the bubbles and the ram pressure of the circularization flow
(vortex).
We emphasize this point: The jets, with the cocoon and bubbles, compress the
dense equatorial gas. This is in contrast with many models which are based
on preexisting equatorial dense gas that collimate the flow.
The second interesting feature is the hot $\sim 1-3\times 10^{6} \K$ gas formed by
the post-shock jet's material.
This gas is a source of X-ray emission, which will be studied in a future paper.

Precessing narrow jets will have the same effect as wide jets in inflating bubbles,
if the precessing angle is large ($\sim 50^\circ$; Sternberg \& Soker 2008).
The same holds for narrow jets bent by the ram pressure of the slow wind, if they are bent
by a large angle.

\section{ANGULAR MOMENTUM}
\label{sec:angular}

We suggest that the back-flowing material discussed above forms the equatorial
large disk found around close binary post-AGB stars.
But where does the required angular momentum come from?
In the binary model for shaping planetary nebulae there are always departures from axisymmetry
(Soker \& Rappaport 2001). The jets are launched by the companion that is rotating around
the center of mass. Two effects cause the jet's axis to be inclined to the $z$ direction
(symmetry axis).
First is the orbital motion.
Let us consider a jet launched by the secondary star at a speed of $400-800 \km \s^{-1}$
with its symmetry axis perpendicular to the equatorial plane in the secondary frame of reference.
With a secondary orbital velocity of $\sim 20-30 \km \s^{-1}$ the jet
axis relative to the nebula will be deflected from the $z$ axis by $\sim 1- 5 ^\circ$.
In addition, the jet will be bent by the ram pressure of the primary stellar wind
(Soker \& Rappaport 2001).
An eccentric orbit, as found in many of these systems, increases the effect of a temporal departure
from axisymmetry.

As a result of the bending, the inflated bubble will not be completely axisymmetric at any given time.
In addition to the ram and thermal pressures toward the equatorial plane, there will be
a small toroidal pressure gradient exerted by the bubble in the direction of orbital motion
as the bent jets rotate around.
This component will be very small, but nonetheless, sufficient to form a Keplerian disk because
the material forming the disk in our model falls back from a large distance of thousands
of AU to hundreds of AU, where the disk forms.
The material escaping the system caries angular momentum with opposite sign, such that
the total angular momentum in the system is conserved.

Our conjecture that this departure from pure axisymmetry can lead to Keplerian disk formation
should be examined with 3D numerical simulations, which is a topic of a future study.
At this point we can only demonstrate the conjecture with the following estimate.
Consider that the thermal pressure gradient and ram pressure accelerate matter to speed
of $v_1 < v_a < v_j$. We scale with $v_a = (v_1 v_j )^{1/2} \sim 50 \km \s^{-1}$,
as we also find in the simualtions.
The matter flowing toward the equatorial plane will be decelerated by the collision
with matter coming from the other side of the equatorial plane, and its motion perpendicular
to the orbital plane will vanish on average.
On the other hand, the toroidal acceleration is in the same direction on both
sides of the equatorial plane.
Let the toroidal acceleration lead to a toroidal speed of $v_t=\chi v_a$ at distance $x_t$
from the center in the equatorial plane.
The bent jets sweep around at a `phase-speed' of $\omega x_t$, where $\omega=2 \pi/P$,
and  $P \sim 1 \year$ is the orbital period.
We take $\chi \sim v_j/\omega x_t$, which for $x_t \sim 10^4 \AU$ and $v_j  \sim 600 \km \s^{-1}$
gives $\chi \sim 0.002$, hence $v_t \sim 0.002 v_a \sim 0.1 \km \s^{-1}$.
The specific angular momentum of the compressed mass is then, for the above values,
$j_t = v_t x_t \sim v_j v_a/\omega \sim 10^3 \km \s^{-1} \AU$.
For a central object mass of $1 M_\odot$ the corresponding Keplerian orbit is
at $x_K \sim 10^3 \AU$.
We conclude that in principle a departure from temporarily axisymmetry, as expected in
binary systems, can lead to non-zero angular momentum of the bound equatorial gas
formed by a back-flowing material.

If the bubbles are inflated by precessing jets, or jets bent by the ram pressure of the slow wind,
then the specific angular momentum of the compressed torus will be even larger.

The viscosity interaction time scale to spread material in geometrically thin disks is
$t_d \sim 0.1 R^2/\alpha_d C_s H$ (Frank et al. 1985), where $R$ is the radius of the disk,
$H \simeq (C_s/v_K) R$ is the vertical size of the disk, $v_K(R)$ is the Keplerian velocity
at $R$, and $\alpha_d \sim 0.01-1$ is the viscosity parameter.
For $H/R \la 0.3$ and a central binary mass of $\sim 1 M_\odot$, we find the
viscosity time to be  $t_d \ga 5000 (R/1000 \AU)^{3/2} /\alpha \year$.
This is not much shorter than the age of post-AGB stars.
In our model, therefore, a disk is built, but it does not reach a complete
equilibrium, and does not settle into a fully thin disk.
We note that this time scale does not allow for angular momentum to be
transport by viscosity from the binary system to the disk.

\section{SUMMARY}
\label{sec:summary}

We propose that the large circumbinary rotating disks found around post-AGB stars result
from slow-wind material that is pushed back to the center
by wide jets that interact with the slow AGB or post-AGB wind.
A large circularization and back-flow, the `cocoon', is formed, which together with the thermal
pressure of the shocked gas compress equatorial gas.
Performing 2D numerical simulations we found that the mass of the bound material
can reach values of $\sim 10^{-4}-10^{-3} M_\odot$, and it flows-back from distances of
$\sim 10^3-10^4 \AU$.

We emphasize that in this paper we don't try to explain all forms of bipolar PNs, and not
all morphological features.
In particular, we are not aiming at explaining nebulae with narrow lobes, which
require narrow jets (Lee \& Sahai 2003; Dennis et al. 2007) and for which the compression of
the material in the equatorial plane might require a different approach (Sahai et al. 2005).
We also did not try to explain expanding dense equatorial torii or small disks having sizes of
$\la 100 \AU$ for which the viscosity time scale is shorter than the age of the system.
We are aiming only at explaining bound rotating material at $\sim 10^3 \AU$ from the central
binary system.

The main ingredients of the proposed model are as follows.
\begin{enumerate}
\item A companion that accrete mass and blow wide jets.
The jets cannot be too fast or too slow. Speeds of $\sim 600 \km \s^{-1}$, as appropriate for
solar-like and somewhat less massive main sequence stars, work the best.
\item During the jet-launching phase, the companion must accrete a large fraction of the
mass lost by the primary AGB or post-AGB star, in particular in the equatorial plane.
Otherwise, the outward flowing wind will destroy the disk. This implies that the companion should
be close to the envelope, but outside the envelope. Either the mass transfer occurs via a
Roche lobe overflow, or the companion resides in the acceleration zone of the primary wind.
\item For the jets to form a large 'cocoon' the jets cannot expand as narrow jets to large distances.
This implies that the jets should have a large opening angle (wide jets),
or be narrow and precess fast or be bent by the ram pressure of the slow wind very close to
the center.
\item Because of ($i$) the orbital motion,
($ii$) the primary stellar wind that hits the jets launched by the secondary star, and/or
($iii$) eccentricity,
the axes of the two jets are not exactly perpendicular to the orbital plane.
The two jets are bent relative to the symmetry axis in the same direction.
This direction changes periodically as the companion orbit the center of mass, and might lead to
small component of toroidal acceleration of the mass flowing back to the center.
With a crude analysis we found that it is enough that this toroidal acceleration component
is $\sim 0.001-0.01$ times the magnitude of the acceleration toward the equatorial plane to
form a large Keplerian disk.
The specific angular momentum of the compressed equatorial matter will be large in cases where
the bubbles are inflated by precessing or bent narrow jets.
Future 3D numerical simulations will have to examine the conditions for the bound material
to posses enough specific angular momentum to form a large Keplerian disk.
\item The viscosity time scale of large disks ($\sim 1000 \AU$) is not much shorter than the
age of these systems. We therefore don't expect large post-AGB disk to be in full
equilibrium.
\end{enumerate}

There are several questions that our study could not answer, and must be addressed before
our proposed flow structure can be accepted as a possible explanation for
the large disks observed
around most post-AGB binary stars with orbital period of $\sim 1~$year.
Firstly, the conditions for the back-flow to occur must be quite common.
Namely, these binary systems must interact in a way that will lead to wide jets
with velocity of $v_j \simeq 400-800 \km \s^{-1}$.
Secondly, the jet interaction with the slow wind forms a large nebula.
We predict that a large nebula exists around each of these systems.
However, in some cases the nebula will be very large, and therefore of low density
and below detection limits.
Thirdly, parameters appropriate for specific nebulae should be
examined. For example, it should be demonstrated that bound mass of
$\sim 0.01 M_\odot$ (Bujarrabal et al. 2003, 2005) can be obtained
to explain the circumbinary disk in the red rectangle; we obtained a mass of
few$\times 10^{-4} M_\odot$ for the parameters we used in several simulations.
This requires an extensive study, as there are many unknown in the process
of jet-launching.

We thank John Blondin for his immense help with the numerical code and for
his comments on the manuscript.
This research was supported by the Asher Space Research Institute.

\end{document}